\documentclass[showpacs,pra,twocolumn]{revtex4}
\usepackage{graphicx}
\usepackage{amsmath}
\newcommand{\ket}[1]{|#1\rangle}
\newcommand{\bra}[1]{\langle#1|}
\newcommand{\Real}{{\textrm{Re}}}

\newcommand{\Tr}{{\textrm{Tr}}} 

\begin{document}
\title{Manifestations of quantum holonomy in interferometry}
\author{Erik Sj\"oqvist$^{1}$\footnote{Electronic address: 
eriks@kvac.uu.se}, 
David Kult$^{1}$\footnote{Electronic address: 
david.kult@kvac.uu.se}, and 
Johan \AA berg$^{2}$\footnote{Electronic address: 
J.Aberg@damtp.cam.ac.uk}} 
\affiliation{$^{1}$Department of Quantum Chemistry, Uppsala University, 
Box 518, Se-751 20 Uppsala, Sweden. \\ 
$^{2}$Centre for Quantum Computation, Department of 
Applied Mathematics and Theoretical Physics, University of Cambridge, 
Wilberforce Road, Cambridge CB3 0WA, United Kingdom.} 
\date{\today}
\begin{abstract} 
Abelian and non-Abelian geometric phases, known as quantum holonomies,
have attracted considerable attention in the past. Here, we show that
it is possible to associate nonequivalent holonomies to discrete
sequences of subspaces in a Hilbert space. We consider two such
holonomies that arise naturally in interferometer settings. For
sequences approximating smooth paths in the base (Grassmann) manifold,
these holonomies both approach the standard holonomy. In the
one-dimensional case the two types of holonomies are Abelian and
coincide with Pancharatnam's geometric phase factor. The theory is
illustrated with a model example of projective measurements involving
angular momentum coherent states.
\end{abstract}
\pacs{03.65.Vf} 
\maketitle
\section{Introduction}
The Abelian geometric phase in the sense of Berry \cite{berry84} and
Pancharatnam \cite{pancharatnam56}, or non-Abelian holonomies in the
sense of Wilczek and Zee \cite{wilczek84} are associated with curves
in a Grassmann manifold \cite{greub73}, i.e., the collection of all
subspaces of a given dimension in a Hilbert space. Such curves may be
realized in adiabatic evolution of a system dependent on external
parameters \cite{berry84,wilczek84} or through a sequence of projective 
filtering measurements of observables \cite{samuel88,anandan89}. In 
these contexts, non-Abelian holonomies arise in cases where the 
parameter dependent Hamiltonian is degenerate and where the measured 
observables have degenerate eigenvalues. The former scenario has 
attracted considerable attention in the literature  
\cite{moody86,tycko87,mead87,zee88,martinez90,arovas98,unanyan99} and has 
recently been shown to be of relevance to robust quantum computation 
\cite{zanardi99,pachos00,duan01,pachos02,faoro03,solinas04,cen04,
fuentes05,sarandy06}. While the latter approach to non-Abelian
holonomies has been discussed in the limit of dense sequences of
projection measurements in Ref. \cite{anandan89}, a detailed analysis
of the genuinely discrete non-Abelian setting, analogous to
Pancharatnam's original discussion \cite{pancharatnam56} of the
Abelian geometric phase in the context of interference of light waves
transmitted by a filtering analyzer, seems still lacking.

In this paper, we examine quantum holonomy in the discrete setting,
and thus complement the study of holonomies in the continuous setting
pursued in Ref.~\cite{kult06}. We show that the discrete setting is
``rich'' in the sense that it admits more than one reasonable type of
holonomy. We demonstrate two distinct holonomies that arise naturally
in this context We shall call these discrete holonomies `direct' and
`iterative'. Although they are nonequivalent, the two types of
holonomies nevertheless approach, in the limit of dense sequences, the
Wilczek-Zee holonomy \cite{wilczek84} for closed paths, as well as its
generalization \cite{kult06} for open paths, which appears to suggest
that the extra richness of the discrete setting disappears in the
continuous limit.  Furthermore, in order to ensure that the direct 
and iterative holonomies are reasonable, we formulate them in
terms of interferometric procedures, thus making them meaningful 
in an operational sense.

The outline of this paper is as follows. In the next section, we
introduce the concepts of direct and iterative holonomies in the
Abelian case followed by their non-Abelian generalizations. We show how
the two holonomies can be associated with the internal degrees of
freedom (e.g., spin) of a particle in an ordinary two-path
interferometer. Section \ref{sec:partial} contains an analysis of the
case where one or several of the adjacent subspaces partially overlap,
leading to the concepts of partial direct and iterative holonomies. An
example involving sequential selections of angular momentum coherent
states is given in Sec. \ref{sec:cs}. The paper ends with the
conclusions.

\section{Holonomy in interferometry} 
Relative phases can be measured in interferometry as shifts in
interference oscillations caused by local manipulations of the
internal states of the interfering particles. In its simplest form,
this can be realized for a pure internal input state $\psi$ that
undergoes a unitary transformation $U$ in one of the interferometer
arms. This results in an interference shift $\arg \bra{\psi} U 
\ket{\psi}$ and visibility $\big| \bra{\psi} U \ket{\psi} \big|$, 
where the former is the Pancharatnam relative phase \cite{pancharatnam56}.  

The above interferometer scenario can be used to develop two different
holonomy concepts that are associated with the geometry of a sequence
of points in a Grassmann manifold, i.e., the set of $K$-dimensional
subspaces of an $N$-dimensional Hilbert space. These concepts we shall
call the direct and iterative holonomies of the sequence. The former
type of holonomy is direct in the sense that the whole operator
sequence representing the points in the Grassmannian is applied to the
internal state in one of the arms of a single interferometer. The
latter type of holonomy is iterative in the sense that it is built up
in several steps, where each step involves an interferometer setup that
depends on the preceding one. For one-dimensional ($K=1$) subspaces,
corresponding to sequences of pure states, the two holonomies are
Abelian phase factors, while for higher dimensional subspaces ($K>1$)
they correspond to non-Abelian unitarities. In the following, we
describe how the two types of quantum holonomies arise in interferometry
in the Abelian and non-Abelian cases.

\subsection{Abelian case}
Let $\psi_1, \ldots, \psi_m$ be a sequence $c$ of pure states with
corresponding one-dimensional projectors $\ket{\psi_1} \bra{\psi_1}$,
$\ldots$, $\ket{\psi_m} \bra{\psi_m}$. We assume that 
$\bra{\psi_{a+1}}\psi_a \rangle \neq 0$, $a=1,\ldots,m-1$, and 
$\bra{\psi_1} \psi_m \rangle \neq 0$. 

Let us first discuss the direct holonomy associated 
with the sequence $c$. Consider particles prepared in the state 
\begin{eqnarray}
\ket{\Psi_0} = \ket{\psi_1} \otimes \frac{1}{\sqrt{2}} 
\big( \ket{0} + \ket{1} \big) , 
\end{eqnarray}
where $\ket{\psi_1}$ is the internal state, and $\ket{0}$ and
$\ket{1}$ represent the two interferometer arms. The internal state is
exposed to the sequence of projection measurements corresponding to
$c$ in the $0$-arm, while a U(1) shift $e^{i\kappa}$ is applied to the
$1$-arm. The filtering measurements correspond to the projection
operators $\pi_a = \ket{\psi_a} \bra{\psi_a} \otimes \ket{0} \bra{0} + 
\hat{1} \otimes \ket{1} \bra{1}$, $a=1,\ldots,m$, where $\hat{1}$ is 
the identity operator on the internal Hilbert space. A 50-50 
beam-splitter yields the (unnormalized) output state 
\begin{eqnarray} 
\ket{\Psi (\kappa)} & = & \frac{1}{2} \Big( \Gamma [c] + 
e^{i\kappa} \hat{1} \Big) \ket{\psi_1} \otimes \ket{0} 
\nonumber \\ 
 & & + \frac{1}{2} \Big( \Gamma [c] -  
e^{i\kappa} \hat{1} \Big) \ket{\psi_1} \otimes \ket{1} ,   
\label{eq:gintabelian}
\end{eqnarray} 
where $\Gamma[c] = \ket{\psi_m} \bra{\psi_m} \ldots \ket{\psi_1} 
\bra{\psi_1}$. The shift of the interference oscillations in the 
$0$-arm produced by varying $\kappa$, is determined by the phase
factor
\begin{eqnarray} 
\gamma_D = \Phi[ \bra{\psi_1} \Gamma [c] \ket{\psi_1} ] ,    
\end{eqnarray}
where $\Phi [z] \equiv z/|z|$ for any nonzero complex 
number $z$. The phase factor $\gamma_D$ is the direct holonomy 
of the sequence $c$.  

The concept of iterative holonomy involves a sequence of
interferometer experiments, each of which being dependent on the
preceding one. Prepare the state
\begin{eqnarray} 
\ket{\Psi_0^{2,1}} = \frac{1}{\sqrt{2}} \Big(  
\ket{\psi_2} \otimes \ket{0} + \ket{\psi_1} \otimes \ket{1} \Big), 
\end{eqnarray}
apply the U(1) phase shift $e^{i\kappa_2}$ to the $0$-arm, and 
let it pass a 50-50 beam-splitter to yield the output state  
\begin{eqnarray} 
\ket{\Psi^{2,1} (\kappa_2)} & = & \frac{1}{2} \Big( e^{i\kappa_2} 
\ket{\psi_2} + \ket{\psi_1} \Big) \otimes \ket{0} 
\nonumber \\ 
 & & + \frac{1}{2} \Big( e^{i\kappa_2} \ket{\psi_2} - \ket{\psi_1} \Big) 
\otimes \ket{1} . 
\label{eq:out1st}
\end{eqnarray}
The resulting intensity $\frac{1}{4} \big\Vert e^{i\kappa_2} 
\ket{\psi_2} + \ket{\psi_1} \big\Vert^2$ in the $0$-arm attains 
its maximum for $e^{i\kappa_2} = e^{i\widetilde{\kappa}_2} = 
\Phi [\bra{\psi_2} \psi_1 \rangle]$. Repeat the procedure but 
with $\ket{\psi_1}$ and $e^{i\kappa_2} \ket{\psi_2}$ in Eq. 
(\ref{eq:out1st}) replaced by $e^{i\widetilde{\kappa}_2} 
\ket{\psi_2}$ and $e^{i\kappa_3} \ket{\psi_3}$, respectively. 
This yields   
\begin{eqnarray} 
\ket{\Psi^{3,2} (\kappa_3)} & = & \frac{1}{2} \Big( e^{i\kappa_3} 
\ket{\psi_3} + e^{i\widetilde{\kappa}_2} \ket{\psi_2} \Big) 
\otimes \ket{0} 
\nonumber \\ 
 & & + \frac{1}{2} \Big( e^{i\kappa_3} \ket{\psi_3} - 
e^{i\widetilde{\kappa}_2} \ket{\psi_2} \Big) \otimes \ket{1} 
\end{eqnarray} 
and the corresponding interference maximum in the $0$-arm for 
$e^{i\kappa_3} = e^{i\widetilde{\kappa}_3} = \Phi [\bra{\psi_3} 
e^{i\widetilde{\kappa}_2} \ket{\psi_2}] = \Phi [\bra{\psi_3} 
\psi_2 \rangle] \Phi [\bra{\psi_2} \psi_1 \rangle]$. 
Continuing in this way up to $\psi_m$ and back to $\psi_1$ results in 
the final phase shift 
\begin{eqnarray} 
e^{i\widetilde{\kappa}_1} = \Phi [\bra{\psi_1} \psi_m \rangle] 
\Phi [\bra{\psi_m} \psi_{m-1} \rangle] \ldots 
\Phi [\bra{\psi_2} \psi_1 \rangle] .
\label{eq:iterativeabelian} 
\end{eqnarray}   
We define $\gamma_I = e^{i\widetilde{\kappa}_1}$ to be the iterative 
holonomy of the sequence $c$. 

Both $\gamma_D$ and $\gamma_I$ are geometric in the sense that 
they are unchanged under the gauge transformations $\ket{\psi_a} 
\rightarrow e^{i\beta_a} \ket{\psi_a}$, $a=1,\ldots ,m$, for arbitrary 
real-valued $\beta_a$. Although operationally different, the direct and
iterative holonomies $\gamma_D$ and $\gamma_I$ are numerically equal. 
Indeed, we have  
\begin{eqnarray} 
\gamma_D & = & \Phi [\bra{\psi_1} \psi_m \rangle \bra{\psi_m} 
\psi_{m-1} \rangle \ldots \bra{\psi_2} \psi_1 \rangle] 
\nonumber \\ 
 & = & \Phi [\bra{\psi_1} \psi_m \rangle] \Phi [\bra{\psi_m} 
\psi_{m-1} \rangle] \ldots \Phi [\bra{\psi_2} \psi_1 \rangle] ,  
\end{eqnarray}
which is $\gamma_I$ according to Eq. (\ref{eq:iterativeabelian}). In fact,
$\gamma_D$ and $\gamma_I$ are both equal to the Pancharatnam geometric 
phase factor \cite{pancharatnam56,ramaseshan86,mukunda93}. 

\subsection{Non-Abelian case}
Consider a sequence $\mathcal{C}$ of discrete points $p_1, p_2, 
\ldots,p_m$ in the Grassmann manifold, now with arbitrary subspace 
dimension $K$. There is a natural bijection between the Grassmann
manifold and the collection of projectors of rank $K$. Thus, we may
associate to $\mathcal{C}$ a sequence $\mathcal{C}'$ of projectors 
$P_1, \ldots, P_m$. We construct the intrinsically geometric 
quantity \cite{kult06}
\begin{eqnarray}
\label{gammadef}
\Gamma [\mathcal{C}] = P_m \ldots P_1 ,
\end{eqnarray}
which is the non-Abelian counterpart to $\Gamma [c]$ in
Eq. (\ref{eq:gintabelian}). Physically, $\Gamma [\mathcal{C}]$ can be
viewed as a sequence of incomplete projective filtering measurements
\cite{anandan89}. Let us introduce a frame $\mathcal{F}_a = \{
\ket{a_k} \}_{k=1}^{K}$ for each subspace $p_a$, $a=1,\ldots,m$. 
The set of frames constitutes a Stiefel manifold, which is a fiber 
bundle \cite{nash83} with the Grassmannian as base manifold and the 
set of $K$-dimensional unitary matrices as fibers. We introduce the 
overlap matrix \cite{mead91,mead92} 
\begin{eqnarray}
(\mathcal{F}_a|\mathcal{F}_b)_{kl} = \bra{a_k} b_l \rangle ,
\end{eqnarray}
which is used in Ref. \cite{kult06} to define holonomy for a
continuous open path in the Grassmannian. The polar decomposition
$\big| (\mathcal{F}_a|
\mathcal{F}_b) \big| \boldsymbol{U}_{a,b}$ of the overlap matrix, 
where $\big| (\mathcal{F}_a|
\mathcal{F}_b) \big| = \sqrt{(\mathcal{F}_a|
\mathcal{F}_b) (\mathcal{F}_b|
\mathcal{F}_a) }$, leads to the definition of
relative phase $\boldsymbol{U}_{a,b}$ as
\begin{eqnarray}
\boldsymbol{U}_{a,b} = \big|(\mathcal{F}_a|
\mathcal{F}_b)\big|^{-1} (\mathcal{F}_a|\mathcal{F}_b)   
\end{eqnarray}
under the assumption that the inverse 
$\big|(\mathcal{F}_a| \mathcal{F}_b)\big|^{-1}$ exists. The 
existence of the inverse is guaranteed if $\big|(\mathcal{F}_a| 
\mathcal{F}_b)\big| >0$, in case of which we say the 
two subspaces $p_a$ and $p_b$ are overlapping \cite{kult06}. 
For overlapping subspaces, the relative phase 
$\boldsymbol{U}_{a,b}$ is a unique unitary matrix.

\begin{figure}[htb]
\includegraphics[width = 8.5cm]{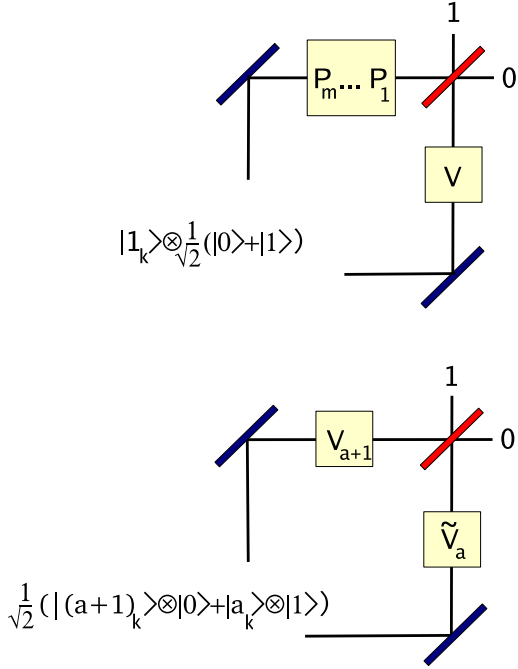}
\caption{\label{fig:fig1} (Color online) Direct (upper panel) 
and iterative (lower panel) holonomy in the interferometer setting. In
the direct scenario, the sequence $P_1, \ldots ,P_m$ of filtering
measurements is applied to the internal state in the $0$-arm (upper
path). In the $1$-arm (lower path), the internal state is exposed to a
unitary operation $V$. The intensity for each orthonormal basis vector
$\ket{1_k}$ of the initial subspace is measured in one of the output
beams. Maximum of the total intensity, defined as the sum over all
$k$, is obtained when $V$ coincide with the direct holonomy of the
sequence. In the iterative scenario, the internal states $\ket{a_k}$
and $\ket{(a+1)_k}$ are exposed to the unitary operations
$\widetilde{V}_a$ and $V_{a+1}$, respectively, in the two
arms. Maximum of the total intensity (sum over $k$) is obtained by
varying $V_{a+1}$ and keeping $\widetilde{V}_a$ fixed. In this way,
the unitary operators $\widetilde{V}_2,
\ldots,\widetilde{V}_m,\widetilde{V}_1$ are given in an iterative
manner, yielding the iterative holonomy as the final unitary 
$\widetilde{V}_1$.}
\end{figure}

We first consider the direct holonomy. A beam of particles is
prepared in an internal state represented by the vector $\ket{1_k} 
\in \mathcal{F}_1$ and divided by a 50-50 beam-splitter, yielding 
the state 
\begin{eqnarray}
\ket{\Psi_k} = \ket{1_k} \otimes \frac{1}{\sqrt{2}}   
\big( \ket{0} + \ket{1} \big).
\end{eqnarray} 
In the $0$-arm the internal state is exposed to the sequence 
$\mathcal{C}'$ of projective filtering measurements, corresponding 
to the action of the projection operators $\Pi_a = P_a \otimes 
\ket{0} \bra{0} + \hat{1} \otimes \ket{1} \bra{1}$, $a=1,\ldots,m$. 
A unitary $V$ is applied to the internal degrees of freedom in the 
other arm. The resulting state pass a 50-50 beam-splitter. 
The output intensity in the $0$-arm reads 
\begin{eqnarray} 
\mathcal{I}_k & = & \frac{1}{4} \Big( 1 + \bra{1_k} 
\Gamma^{\dagger} [\mathcal{C}] \Gamma [\mathcal{C}] 
\ket{1_k} \Big) 
\nonumber \\ 
 & & + \frac{1}{2} \Real \big[ \boldsymbol{V}^{\dagger} 
\boldsymbol{D} \big]_{kk} ,   
\end{eqnarray}
where $\big[ \boldsymbol{V} \big]_{kl} = \bra{1_k} V \ket{1_l}$ 
is a unitary $K\times K$ matrix and we have introduced the matrix 
product 
\begin{eqnarray}
\boldsymbol{D} & = & (\mathcal{F}_1|\mathcal{F}_m)
(\mathcal{F}_m|\mathcal{F}_{m-1}) \ldots 
(\mathcal{F}_2|\mathcal{F}_1) .
\end{eqnarray}
Summing over all $k$ yields the total intensity 
\begin{eqnarray} 
\mathcal{I}_{\textrm{tot}} & = & \sum_{k=1}^K \mathcal{I}_k =  
\frac{1}{4} \Big( K + \Tr \big( \Gamma^{\dagger} 
[\mathcal{C}] \Gamma [\mathcal{C}] \big) \Big)   
\nonumber \\ 
 & & + \frac{1}{2} \Real \Tr \big( \boldsymbol{V}^{\dagger} 
\boldsymbol{D} \big) .   
\end{eqnarray}
Under the assumption that $\big| \boldsymbol{D} \big|^{-1}$ exists, 
the total intensity attains its maximum when 
\begin{eqnarray} 
\boldsymbol{V} = \boldsymbol{U}_D \equiv 
\big| \boldsymbol{D} \big|^{-1} \boldsymbol{D} . 
\end{eqnarray}
The unitary matrix $\boldsymbol{U}_D$ is the direct holonomy 
associated with the sequence $\mathcal{C}$ as measured in the 
interferometry setup shown in the upper panel of Fig. \ref{fig:fig1}.

Next, we consider the iterative holonomy, which, as in the Abelian
case, involves the performance of a sequence of interferometry 
experiments. Suppose all adjacent subspaces of the extended sequence 
$p_1,\ldots,p_m,p_1$ are overlapping. Prepare the state
\begin{eqnarray}
\ket{\Psi_k^{2,1}} = \frac{1}{\sqrt{2}} \Big( V_2 \ket{2_k} 
\otimes \ket{0} + \ket{1_k} \otimes \ket{1} \Big) , 
\end{eqnarray}
where $V_2 P_2 V_2^{\dagger} = P_2$. A 50-50 beam-splitter yields 
the output intensity in the $0$-arm as 
\begin{eqnarray}
\mathcal{I}_k^{2,1} & = & \frac{1}{4} 
\big\Vert \ket{1_k} + V_2 \ket{2_k} \big\Vert^2
\nonumber \\ 
 & = &  \frac{1}{2} \Big( 1 + \Real \big[ 
(\mathcal{F}_1|\mathcal{F}_2) \boldsymbol{V}_2 \big]_{kk} \Big) ,
\end{eqnarray} 
where $\big[ \boldsymbol{V}_2 \big]_{kl} = \bra{2_k}V_2\ket{2_l}$ 
is a unitary $K \times K$ matrix. Summing over $k$ yields the 
total intensity 
\begin{eqnarray}
\mathcal{I}_{\textrm{tot}}^{2,1} = 
\sum_{k=1}^K \mathcal{I}_k^{2,1} = \frac{1}{2} \Big( K + 
\Real \Tr \big[ (\mathcal{F}_1|\mathcal{F}_2) 
\boldsymbol{V}_2 \big] \Big) ,
\end{eqnarray} 
which attains its maximum for $\boldsymbol{V}_2 = 
\widetilde{\boldsymbol{V}}_2 = \boldsymbol{U}_{2,1}$. 
In the next step, prepare 
\begin{eqnarray}
\ket{\Psi_k^{3,2}} = \frac{1}{\sqrt{2}} \Big( V_3 \ket{3_k} \otimes \ket{0} + 
\widetilde{V}_2 \ket{2_k} \otimes \ket{1} \Big) ,
\end{eqnarray} 
where $V_3 P_3 V_3^{\dagger} = P_3$ and $\bra{2_k} \widetilde{V}_2 
\ket{2_l} = \big[ \widetilde{\boldsymbol{V}}_2 \big]_{kl}$. The two beams 
are made to interfere by a 50-50 beam-splitter. Adding the resulting
output intensities yields
\begin{eqnarray}
\mathcal{I}_{\textrm{tot}}^{3,2} & = &  
\frac{1}{2} \Big( 1 + \Real 
\Tr \big[ \boldsymbol{U}_{2,1}^{\dagger}
\big| (\mathcal{F}_2|\mathcal{F}_3) \big| 
\boldsymbol{U}_{2,1} 
\nonumber \\ 
 & & \times \big( \boldsymbol{U}_{3,2} 
\boldsymbol{U}_{2,1} \big)^{\dagger} \boldsymbol{V}_3 \big] \Big) , 
\end{eqnarray} 
which is maximal for $\boldsymbol{V}_3 = \widetilde{\boldsymbol{V}}_3 = 
\boldsymbol{U}_{3,2} \boldsymbol{U}_{2,1}$. By continuing in this way up 
to $P_m$ and back to $P_1$, we obtain the final result 
\begin{eqnarray} 
\label{UI}
\widetilde{\boldsymbol{V}}_1 = \boldsymbol{U}_I \equiv 
\boldsymbol{U}_{1,m} \boldsymbol{U}_{m,m-1} \ldots 
\boldsymbol{U}_{2,1} . 
\end{eqnarray}
The unitary matrix $\boldsymbol{U}_I$ is the iterative holonomy
associated with $\mathcal{C}$. The interferometer setting giving rise
to the iterative holonomy is illustrated in the lower panel of
Fig. \ref{fig:fig1}.

Under the change of frames 
\begin{eqnarray}
\mathcal{F}_a \rightarrow 
\Big\{ \sum_{k=1}^K \ket{a_k} \big[ \boldsymbol{W}_a \big]_{k,l} 
\Big\}_{l=1}^K, \ a = 1,\ldots ,m,
\label{eq:change}
\end{eqnarray}
$\boldsymbol{W}_a$ being unitary matrices, we have 
\begin{eqnarray} 
(\mathcal{F}_{a+1}|\mathcal{F}_a) & \rightarrow & 
\boldsymbol{W}_{a+1}^{\dagger} (\mathcal{F}_{a+1}|\mathcal{F}_a) 
\boldsymbol{W}_a , 
\nonumber \\ 
\boldsymbol{U}_{a+1,a} & \rightarrow & 
\boldsymbol{W}_{a+1}^{\dagger} \boldsymbol{U}_{a+1,a} 
\boldsymbol{W}_a .
\label{eq:gt}
\end{eqnarray} 
Such a change of frames can be seen as a gauge transformation, 
i.e., a motion along the fiber over each of the points 
$p_1,\ldots,p_m$ in the Grassmannian. From Eq. (\ref{eq:gt}) 
\begin{eqnarray}
\boldsymbol{U}_D & \rightarrow & \boldsymbol{W}_1^{\dagger}
\boldsymbol{U}_D \boldsymbol{W}_1 , 
\nonumber \\ 
\boldsymbol{U}_I & \rightarrow & \boldsymbol{W}_1^{\dagger} 
\boldsymbol{U}_I \boldsymbol{W}_1, 
\end{eqnarray} 
i.e., the direct and iterative holonomies transform unitarily 
(gauge covariantly) under change of frames.

The unitary matrices $\boldsymbol{U}_D$ and $\boldsymbol{U}_I$ are the
non-Abelian generalizations of $\gamma_D$ and $\gamma_I$,
respectively. However, while $\gamma_D = \gamma_I$, we have
$\boldsymbol{U}_D \neq \boldsymbol{U}_I$ in general. There are 
situations, though, where the two approaches give
the same result, e.g., for continuous paths in the Grassmannian. 
This follows from the fact that for a smooth choice of 
$\mathcal{F}_s = \{ \ket{a_k (s)} \}_{k=1}^K$, we have 
$\big| (\mathcal{F}_{s+\delta s}|\mathcal{F}_s) \big| =
\boldsymbol{1} + O(\delta s^2)$, $\boldsymbol{1}$ being the 
$K$-dimensional identity matrix. Thus, for $s\in [0,1]$ 
we obtain 
\begin{eqnarray}
\boldsymbol{D} & = & (\mathcal{F}_0|\mathcal{F}_1) 
\big( \boldsymbol{1} + O(\delta s^2) \big)
\boldsymbol{U}_{1,1-\delta s} \ldots 
\nonumber \\ 
 & & \times \big( \boldsymbol{1} + O(\delta s^2) \big)
\boldsymbol{U}_{\delta s,0}
\nonumber \\
 & = & (\mathcal{F}_0|\mathcal{F}_1) 
\boldsymbol{U}_{1,1-\delta s} \ldots
\boldsymbol{U}_{\delta s,0} + O(\delta s) , 
\end{eqnarray}
where the correction term is of order $O(\delta s)$ since it contains 
$\delta s^{-1}$ terms. By using the assumption that $\big|
(\mathcal{F}_0|\mathcal{F}_1) \big|$ is invertible and the fact that
$\boldsymbol{U}_{1,1-\delta s} \ldots \boldsymbol{U}_{\delta s,0}$ is
guaranteed to be unitary for sufficiently small $\delta s$, we have $\big|
\boldsymbol{D} \big| = \big| (\mathcal{F}_0|\mathcal{F}_1)
\big| + O(\delta s)$ and $\big| \boldsymbol{D} \big|^{-1} = 
\big| (\mathcal{F}_0|\mathcal{F}_1) \big|^{-1} + O(\delta s)$. 
It follows that  
\begin{eqnarray} 
\boldsymbol{U}_D & = & \boldsymbol{U}_{0,1} 
\boldsymbol{U}_{1,1-\delta s} \ldots \boldsymbol{U}_{\delta s,0} + 
O(\delta s) 
\nonumber \\ 
 & = & \boldsymbol{U}_I + O(\delta s)
\end{eqnarray}
since $\big| (\mathcal{F}_0|\mathcal{F}_1) \big|^{-1} 
(\mathcal{F}_0|\mathcal{F}_1) = \boldsymbol{U}_{0,1}$.   
Thus, in the $\delta s \rightarrow 0$ limit, we obtain   
\begin{eqnarray}
\boldsymbol{U}_D & = & \boldsymbol{U}_I = 
\boldsymbol{U}_{0,1} {\bf P} e^{\int_0^1 \boldsymbol{A}(s)ds}
\end{eqnarray} 
with $[\boldsymbol{A}(s)]_{kl}=\langle
\dot{a}_k(s)\ket{a_l(s)}$.  In other words, in the 
continuous path limit, the direct and iterative holonomies 
are equal to the Wilczek-Zee holonomy \cite{wilczek84} for 
closed paths (for which $\boldsymbol{U}_{0,1} = \boldsymbol{1}$), 
as well as its generalization \cite{kult06} for open paths.

We finish this section by pointing out a relation between the above
iterative holonomy and the Uhlmann holonomy \cite{uhlmann86} applied
to a special class of density operators 
\cite{remark1,sjoqvist00,tong04,slater02,ericsson03,shi05,rezakhani06}. 
This class
consists of normalized rank $K$ projectors, and we consider sequences
$\frac{1}{K} P_1,\ldots,\frac{1}{K} P_m,\frac{1}{K} P_1$ of such
density operators. If all the adjacent subspaces are overlapping, this
is a sufficient condition for these density operators to constitute an
admissible sequence \cite{uhlmann86}, for which the Uhlmann holonomy
$U_{\textrm{uhl}}$ reads
\begin{eqnarray}
\label{Uuhl}
U_{\textrm{uhl}} = \widetilde{U}_{1,m} \widetilde{U}_{m,m-1} \ldots 
\widetilde{U}_{2,1} ,   
\end{eqnarray}
where $\widetilde{U}_{a+1,a}$, $a=1,\ldots,m-1$, and 
$\widetilde{U}_{1,m}$ are the 
partial isometry parts of $P_{a+1} P_a$ and $P_1 P_m$, respectively 
\cite{remark2}.
The overlap matrices can be written
\begin{eqnarray} 
\label{fvnz}
(\mathcal{F}_{a+1}|\mathcal{F}_{a})_{kl} & = &  
\sum_n \bra{(a+1)_k} \big| P_{a+1} P_a \big| \ket{(a+1)_n} 
\nonumber \\ 
 & & \times \bra{(a+1)_n} \widetilde{U}_{a+1,a} \ket{a_l},  
\end{eqnarray}
where $\big| P_{a+1} P_a \big|$ is the positive part of $P_{a+1}
P_a$. One can write $(\mathcal{F}_{1}|\mathcal{F}_{m})$
similarly. From Eq.~(\ref{fvnz}) it follows that
$[\boldsymbol{U}_{a+1,a}]_{nl} = \bra{(a+1)_n} \widetilde{U}_{a+1,a}
\ket{a_l}$. By combining this with Eqs.~(\ref{UI}) and (\ref{Uuhl}),
and using that $\widetilde{U}_{a+1,a} = P_{a+1}
\widetilde{U}_{a+1,a}$, we find 
\begin{equation}
[\boldsymbol{U}_I]_{kl} = \langle 1_{k}|U_{\textrm{uhl}}|1_{l}\rangle,
\end{equation}
for admissible sequences of density operators that are proportional 
to projectors \cite{remark3,aberg06}. 

\section{Partial holonomies} 
\label{sec:partial}
If at least one pair of adjacent states in the extended sequence
$\psi_1,\ldots,\psi_m,\psi_1$ are orthogonal, then the corresponding
holonomies $\gamma_D$ and $\gamma_I$ are undefined. Similarly,
$\boldsymbol{U}_D$ and $\boldsymbol{U}_I$ are undefined if any of the
adjacent pairs of subspaces are orthogonal. However, the non-Abelian
case includes partially defined holonomies, when the number of nonzero
eigenvalues of the positive part of $\boldsymbol{U}_{1,m}
\boldsymbol{U}_{m,m-1} \ldots \boldsymbol{U}_{2,1} \equiv \boldsymbol{I}$ 
or $\boldsymbol{D}$ is greater than zero but less than the subspace
dimension $K$.  This occurs when at least one pair of adjacent
subspaces is partially overlapping, which results in a nonunique
unitary part of the overlap matrix. To remove this nonuniqueness, one
may use the Moore-Penrose (MP) pseudo inverse \cite{lancaster85},
denoted as $\ominus$, to introduce a well-defined concept of relative
phase. Let $\mathcal{F}_a$ and $\mathcal{F}_b$ be two frames of two
partially overlapping subspaces $p_a$ and $p_b$. Then the MP pseudo
inverse is obtained by inverting the nonzero eigenvalues of
$\big|(\mathcal{F}_a|
\mathcal{F}_b)\big|$ in its spectral decomposition. We define
\begin{eqnarray}
\boldsymbol{U}_{a,b} = \big|(\mathcal{F}_a|
\mathcal{F}_b)\big|^{\ominus} (\mathcal{F}_a|\mathcal{F}_b)
\end{eqnarray}
as the relative phase between the two frames. The matrix
$\boldsymbol{U}_{a,b}$ is a unique partial isometry.

In Ref. \cite{kult06}, the relative phase between frames of partially
overlapping subspaces was used to introduce a concept of partial
holonomy of continuous open paths in the Grassmannian. Here, we
develop the corresponding concepts for the discrete sequence
$\mathcal{C}$.

For the direct holonomy to be (totally) defined it is a necessary and
sufficient condition that all the adjacent subspaces (in the extended
sequence) are overlapping \cite{remark4}.  Thus, if there is partially
overlapping subspaces in the sequence, and if there is at least one
nonzero eigenvalue of $\big| \boldsymbol{D} \big|$, then
the MP pseudo inverse yields
\begin{eqnarray}
\boldsymbol{U}_D =
\big| \boldsymbol{D} \big|^{\ominus} \boldsymbol{D},
\label{eq:partialdirect}
\end{eqnarray}
which we define to be the partial direct holonomy.

In the iterative case, we again find that the holonomy becomes
partial or undefined if at least one pair of adjacent subspaces in the
sequence $\mathcal{C}$ is partially overlapping. For such cases, 
$\boldsymbol{I}$ is not unitary since at least one of the matrices
$\boldsymbol{U}_{2,1},\ldots , \boldsymbol{U}_{1,m}$ is a partial
isometry. If $\big| \boldsymbol{I} \big|$ has at least one nonzero
eigenvalue, we define the partial isometry part of
$\boldsymbol{I}$, i.e.,
\begin{eqnarray}
\boldsymbol{U}_I = \big| \boldsymbol{I} \big|^{\ominus} \boldsymbol{I},
\label{eq:partialiterative}
\end{eqnarray}
to be the partial iterative holonomy associated with $\mathcal{C}$.

Let us discuss how the partial holonomies behave under 
gauge transformations. From Eq. (\ref{eq:gt}) we obtain   
\begin{eqnarray}
\boldsymbol{D} & \rightarrow & \boldsymbol{W}_1^{\dagger}
\boldsymbol{D} \boldsymbol{W}_1 \Rightarrow 
\big| \boldsymbol{D} \big| \rightarrow \boldsymbol{W}_1^{\dagger}
\big| \boldsymbol{D} \big| \boldsymbol{W}_1 , 
\nonumber \\ 
\boldsymbol{I} & \rightarrow & \boldsymbol{W}_1^{\dagger} 
\boldsymbol{I} \boldsymbol{W}_1  \Rightarrow 
\big| \boldsymbol{I} \big| \rightarrow \boldsymbol{W}_1^{\dagger}
\big| \boldsymbol{I} \big| \boldsymbol{W}_1 
\label{eq:partialgt1}    
\end{eqnarray} 
under the change of frames in Eq. (\ref{eq:change}). 
Furthermore, for any matrix $\boldsymbol{X}$ and unitary matrices 
$\boldsymbol{U}$ and $\boldsymbol{V}$ we have $(\boldsymbol{U} 
\boldsymbol{X} \boldsymbol{V})^{\ominus} = \boldsymbol{V}^{\dagger} 
\boldsymbol{X}^{\ominus} \boldsymbol{U}^{\dagger}$ (see, e.g., 
p. 434 in Ref. \cite{lancaster85}). Thus,   
\begin{eqnarray}
\big| \boldsymbol{D} \big|^{\ominus} & \rightarrow & 
\boldsymbol{W}_1^{\dagger}
\big| \boldsymbol{D} \big|^{\ominus} \boldsymbol{W}_1 , 
\nonumber \\ 
\big| \boldsymbol{I} \big|^{\ominus} & \rightarrow & 
\boldsymbol{W}_1^{\dagger}
\big| \boldsymbol{I} \big|^{\ominus} \boldsymbol{W}_1 . 
\label{eq:partialgt2}
\end{eqnarray} 
By combining Eqs. (\ref{eq:partialgt1}) and (\ref{eq:partialgt2}),  
it follows that the direct and iterative holonomies transform 
unitarily (gauge covariantly) also in the partial case.  

We prove that the two partial holonomies in Eqs. 
(\ref{eq:partialdirect}) and (\ref{eq:partialiterative}) coincide
with that of Ref. \cite{kult06} in the continuous path limit. To do
this, we consider the smooth choice $\mathcal{F}_s = 
\{ \ket{a_k (s)} \}_{k=1}^K$ and note that for sufficiently 
small $\delta s$, the two holonomies become partial only if $\big|
(\mathcal{F}_0|\mathcal{F}_1) \big|$ is not invertible. In such a 
case, $\big| \boldsymbol{D} \big|^{\ominus} = \big| (\mathcal{F}_0| 
\mathcal{F}_1) \big|^{\ominus} + O(\delta s)$ and $\boldsymbol{U}_I = 
\boldsymbol{U}_{0,1} \boldsymbol{U}_{1,1-\delta s} \ldots 
\boldsymbol{U}_{\delta s,0}$, where $\boldsymbol{U}_{0,1}$ is 
a partial isometry and $\boldsymbol{U}_{1,1-\delta s} \ldots 
\boldsymbol{U}_{\delta s,0}$ is unitary. It follows that 
\begin{eqnarray}
\boldsymbol{U}_D & = & \boldsymbol{U}_I = 
\boldsymbol{U}_{0,1} {\bf P} e^{\int_0^1 \boldsymbol{A}(s)ds} 
\end{eqnarray} 
in the $\delta s \rightarrow 0$ limit, which is the partial holonomy   
of Ref. \cite{kult06}.

\section{Angular momentum coherent states}
\label{sec:cs}
Consider a particle carrying an angular momentum $j$, $j \geq 1$. 
Let $J_{{\bf n}_a}$ be the angular momentum
component in the direction ${\bf n}_a$ characterized by the 
spherical polar angles $\theta_a,\phi_a$, i.e., ${\bf n}_a = (\sin
\theta_a \cos \phi_a, \sin \theta_a \sin \phi_a, \cos
\theta_a)$. Let $\{ \ket{\mu} \}_{\mu=-j}^j$ be the eigenbasis of
$J_z$. Consider a sequence of filtering measurements of $J_{{\bf
n}_a}^2$, $a=1,\ldots,m$, each of which selects the maximal angular
momentum projection quantum numbers $\mu = \pm j$ (angular momentum 
coherent states \cite{peres95}). The selection corresponds to the 
two-dimensional projection operators $P_{{\bf n}_a} = \ket{j;{\bf n}_a} 
\bra{j;{\bf n}_a} + \ket{-j;{\bf n}_a} \bra{-j;{\bf n}_a}$, 
$a=1,\ldots,m$, where $\ket{\pm j;{\bf n}_a}$ are eigenvectors of
$J_{{\bf n}_a}$. The use of angular momentum coherent states
simplifies the subsequent calculation since $\ket{j;{\bf n}_a}$ can be
viewed as a product state of $2j$ copies of the spin-$\frac{1}{2}$
state $\ket{\frac{1}{2};{\bf n}_a}$, and $\ket{-j;{\bf n}_a}$
similarly as $2j$ copies of $\ket{-\frac{1}{2};{\bf n}_a}$.

Now, let ($\hbar = 1$ from now on)
\begin{eqnarray}
\mathcal{F} (\theta_a,\phi_a) =
\{ e^{-i\phi_a J_z} e^{-i\theta_a J_y} \ket{\pm j} \} .
\label{eq:frame}
\end{eqnarray}
For this choice of frames, the overlap matrix takes the form  
\begin{equation}
( \mathcal{F} (\theta_a,\phi_a) | \mathcal{F} (\theta_b,\phi_b) ) =
\left(
\begin{array}{ll}
R(a,b) & S(a,b) \\
(-1)^{2j} S(a,b)^{\ast} & R(a,b)^{\ast}
\end{array}
\right) ,
\label{eq:overlap}
\end{equation}
where
\begin{eqnarray}
R(a,b) & = &
\left[ \cos \left( \frac{\theta_a - \theta_b}{2} \right)
\cos \left( \frac{\phi_a - \phi_b}{2} \right) \right.
\nonumber \\
 & & \left. + i \cos \left( \frac{\theta_a +
\theta_b}{2} \right) \sin \left( \frac{\phi_a - \phi_b}{2} \right)
\right]^{2j} ,
\nonumber \\
S(a,b) & = &
\left[ \sin \left( \frac{\theta_a - \theta_b}{2} \right)
\cos \left( \frac{\phi_a - \phi_b}{2} \right) \right.
\nonumber \\
 & & \left. - i \sin \left( \frac{\theta_a +
\theta_b}{2} \right) \sin \left( \frac{\phi_a - \phi_b}{2} \right)
\right]^{2j} .
\label{eq:cd}
\end{eqnarray}
We notice that $\sqrt[j]{|R(a,b)|} + \sqrt[j]{|S(a,b)|} = 1$, i.e.,
the overlap matrix cannot vanish for this system.

If $j$ is a half-odd integer, then
\begin{eqnarray}
 & & ( \mathcal{F} (\theta_a,\phi_a) | \mathcal{F} (\theta_b,\phi_b) ) 
\nonumber \\ 
 & & = \sqrt{|R(a,b)|^2 + |S(a,b)|^2} \boldsymbol{U}_{a,b} ,
\end{eqnarray}
where $\boldsymbol{U}_{a,b}$ is a unique unitary matrix. 
It follows that the direct and iterative holonomies are identical.

When $j$ is an integer, the overlap matrix in Eq. (\ref{eq:overlap})
may have a nontrivial positive part. This implies that the two types 
of holonomies may be different. To illustrate this, consider the sequence
of directions ${\bf n}_1, {\bf n}_2, {\bf n}_3, {\bf n}_4$
characterized by the polar angles $(\theta_0,\phi_0)$,
$(\theta_1,\phi_0)$, $(\theta_1,\phi_1)$, $(\theta_0,\phi_1)$,
respectively.  Assume that the first and third overlap matrices are
degenerate. This happens for $| \theta_1-\theta_0 | = \pi /2$, which
yields $(\mathcal{F}(\theta_1,\phi_0))| \mathcal{F}(\theta_0,\phi_0))
= 2^{1-j} \boldsymbol{U}_{2,1}$ and $(\mathcal{F}(\theta_0,\phi_1))|
\mathcal{F}(\theta_1,\phi_1)) = 2^{1-j} \boldsymbol{U}_{4,3}$.  
Here, $\boldsymbol{U}_{2,1} = \boldsymbol{U}_{4,3} = 
\frac{1}{2} \big( \boldsymbol{1} + 
\boldsymbol{\sigma}_x \big)$, where $\boldsymbol{1}$ and 
$\boldsymbol{\sigma}_x$ are the $2\times 2$ identity and Pauli$-X$ 
matrices, respectively. We furthermore assume that $|S(3,2)| > |R(3,2)|$ 
and $|S(1,4)| > |R(1,4)|$ for which a polar decomposition yields the 
unitary matrices $\boldsymbol{U}_{3,2} = e^{ij \chi_1  
\boldsymbol{\sigma}_z}$ and $\boldsymbol{U}_{1,4} = e^{-ij \chi_0 
\boldsymbol{\sigma}_z}$, respectively, $\boldsymbol{\sigma}_z$ being 
the Pauli-$Z$ matrix. Here, $\chi_k = 2 \arctan \big[ \cos 
\theta_k \tan \big( \Delta \phi /2 \big) \big]$, $k=0,1$, where 
$\Delta \phi = \phi_1 - \phi_0$. We obtain the partial holonomies 
\begin{eqnarray}
\boldsymbol{U}_D & = & \frac{q_D}{|q_D|} \frac{1}{2}
\left(
\begin{array}{ll}
e^{-i\eta_0} & e^{-i\eta_0} \\ e^{i\eta_0} & e^{i\eta_0}
\end{array}
\right) ,
\nonumber \\
\boldsymbol{U}_I & = & \frac{q_I}{|q_I|} \frac{1}{2}
\left(
\begin{array}{ll}
e^{-ij\chi_0} & e^{-ij\chi_0} \\
e^{ij\chi_0} & e^{ij\chi_0}
\end{array}
\right) ,
\end{eqnarray}
where
\begin{widetext}
\begin{eqnarray}
\eta_0 & = & -\arctan \left[ \frac{\left( 1-\sin^2 \theta_0 \sin^2 
\frac{\Delta \phi}{2} \right)^j \sin (j\chi_0)}{\left( 1- 
\sin^2 \theta_0 \sin^2 \frac{\Delta \phi}{2} \right)^j \cos (j\chi_0) 
+(-1)^j \left( \sin^2 \theta_0 \sin^2
\frac{\Delta \phi}{2} \right)^j} \right] , 
\nonumber \\ 
q_D & = & \left( 1-\sin^2 \theta_1 \sin^2 \frac{\Delta \phi}{2} \right)^j
\cos (j\chi_1) +(-1)^j \left( \sin^2 \theta_1 \sin^2
\frac{\Delta \phi}{2} \right)^j ,
\nonumber \\
q_I & = & \cos (j\chi_1) .
\end{eqnarray}
\end{widetext}
It follows that the direct and iterative holonomies differ unless $q_D$ 
and $q_I$ have the same sign and $\eta_0 = j\chi_0$. The latter 
happens only if $\sin^2 \theta_0 \sin^2 (\Delta \phi /2) = 0$. Note 
that $\boldsymbol{U}_D$ ($\boldsymbol{U}_I$) is undefined if $q_D=0$ 
($q_I=0$). 

\section{Conclusions}
Corresponding to a sequence in the Grassmann manifold of
$K$-dimensional subspaces in an $N$-dimensional Hilbert space, we have
defined two holonomies, which both are gauge covariant in the sense
that they transform unitarily under the change of frames in the
subspaces. Interferometer settings that give rise to the two
holonomies have been delineated. In the non-Abelian case these two
holonomies are generically nonequivalent. In the case of
one-dimensional subspaces, however, both the holonomies reduce to the
standard Pancharatnam phase. Moreover, we have shown that in the limit
where the sequences form a continuous and smooth path in the Grassmann
manifold, the two discrete holonomies coincide with the Wilczek-Zee
holonomy \cite{wilczek84} in the case of closed paths, and its
generalized noncyclic version \cite{kult06}, in the open-path
case. It is an interesting question whether there exist other
discrete holonomies, distinct from the two considered here, that also
converge to the standard holonomy in the limit of smooth curves, and
if so, if those can be implemented interferometrically.

\vskip 0.3 cm 
E.S. acknowledges financial support from the Swedish Research 
Council. J.{\AA}. wishes to thank the Swedish Research Council for 
financial support and the Centre for Quantum Computation at DAMTP, 
Cambridge, for hospitality.


\begin{thebibliography}{99}
\bibitem{remark3} Although the Uhlmann holonomy for the density 
operators $\frac{1}{K} P_1,\ldots,\frac{1}{K}P_m,\frac{1}{K}P_1$
coincides with the iterative holonomy for $P_1,\ldots,P_m,P_1$, no
such relation exists for general sequences of density operators. 
However, this does not rule out the possibility of an interferometric 
interpretation of the Uhlmann holonomy for general sequences. 
Indeed, such an interpretation has been provided in Ref. 
\cite{aberg06}.
\end{thebibliography}
\end{document}